\documentstyle[aps,prb]{revtex}
\draft

\input epsf
\input{epsf.sty}
\begin{document}
\twocolumn[

\hsize\textwidth\columnwidth\hsize
\csname@twocolumnfalse\endcsname

\title{Entropy, vortex interactions and the phase diagram of heavy-ion 
irradiated Bi$_{2}$Sr$_{2}$CaCu$_{2}$O$_{8+\delta}$}

\author{C.J. van der Beek and M. Konczykowski}
\address{Laboratoire des Solides Irradi\'{e}s, C.N.R.S. U.M.R. 7642, Ecole Polytechnique, 91128
         Palaiseau, France }
		 
\author{R.J. Drost and P.H. Kes}
\address{Kamerlingh Onnes Laboratorium, Leiden University, P.O. 
Box 9506, 2300 RA Leiden, the Netherlands}

\author{N.Chikumoto}
\address{Superconductivity Research Laboratory, ISTEC, Minato-ku, 
         Tokyo 105, Japan}
		 
\author{S. Bouffard}
\address{Centre Interdisciplinaire de Recherche avec les Ions 
Lasers (C.I.R.I.L.), B.P. 5133, 14040 Caen Cedex, France}

\date{\today}
\maketitle

\begin{abstract}
Dynamic and thermodynamic magnetization experiments on 
heavy-ion irradiated single crystalline 
Bi$_{2}$Sr$_{2}$CaCu$_{2}$O$_{8+\delta}$ are correlated in order to 
clarify the nature of the mixed state phase diagram. It is shown 
that whereas the entropy contribution to the free energy in the London 
regime plays a minor role in unirradiated crystals and irradiated crystals at fields 
close to or above the matching field $B_{\phi}$, it becomes very important 
at low fields in irradiated crystals with high $B_{\phi}$. The direct 
determination of the entropy contribution to the free energy from the 
reversible magnetization allows one to determine not only the correct values of 
the pinning energy, but also to extract quite detailed information on 
pancake vortex alignment. The characteristic field $H_{int} \sim 
\frac{1}{6} B_{\phi}$ at which 
intervortex repulsion begins to determine the vortex arrangement and 
the reversible magnetization is shown to coincide with a sharp 
increase in the irreversibility field $H_{irr}(T)$ and with the recoupling transition found 
in Josephson Plasma Resonance. Above $H_{int}$, the repulsive 
interaction between vortices cause both the vortex mobility to 
decrease and pancake alignement to increase. At higher fields $\gtrsim 
\frac{1}{3}B_{\phi}\gg B_{c1}$, free vortices outnumber those that are trapped on a 
columnar defect. This causes the decrease of $c$-axis resistivity and a second crossover 
of the irreversibility field, to a regime where it is determined by plastic 
creep.

\end{abstract}
]
\narrowtext

\section{Introduction}
The remarkable influence of heavy ion--irradiation induced amorphous 
columnar defects on the mixed state phase-diagram and transport properties of
layered superconductors, and the interplay between the pinning of 
vortices and coupling between the layers, has provoked a great deal of interest.
\cite{Hardy92,Konczykowski93,Hardy93,Klein94I,Konczykowski95,vdBeek95I,Zech95,Zech96,Seow96,Doyle96,Sato97,Kosugi97,Chikumoto98}
The effect of the ion tracks is twofold. First, the destruction 
of superconductivity within the columnar defect 
core, the radius of which is comparable to the superconducting 
coherence length $\xi$, leads to a large reduction of the mixed state 
free energy \cite{Wahl95,vdBeek96II,QiangLi96,Drost98,Drost99} arising from the pinning of
the two--dimensional pancake vortices in the superconducting layers\cite{Kes90,Clem91I} 
by the columns. The reduction in free energy due to pinning is readily measurable 
because it entails a significant reduction of the reversible 
magnetization $M_{rev}$ at fields well below the matching field $B_{\phi} = 
\Phi_{0}n_{d}$, and a nonmonotonicity of $\partial M_{rev}/\partial \ln B$ as function of $B$ ($\Phi_{0} = 
h/2e$ is the flux quantum and $n_{d}$ is the density of column 
intersections with the layers).\cite{Wahl95,vdBeek96II,QiangLi96} 
The coincidence of the radius of the columnar defect core
with that of the vortex core implies a large pinning force\cite{Nelson92} 
and concommittantly large values of the critical current 
density,\cite{Konczykowski95} as well as a large increase of the field $H_{irr}(T)$  
below which irreversible magnetic response is measured.\cite{Konczykowski95} 
Second, the linear geometry of the ion tracks results in the effective 
alignment of the pancake vortices along the tracks, and the 
re-establishment of interlayer superconducting phase coherence at 
temperatures and fields where the unirradiated material behaves as a 
stack of nearly decoupled superconducting 
layers.\cite{Klein94I,Zech95,Iye91,Raffy91,Martinez92,Matsuda95} The pancake vortex
alignment is manifest in the angular dependence of the transport 
properties of the irradiated superconductor,\cite{Klein94I,vdBeek95I} as well as 
in measurements of the $c$-axis conductivity (perpendicular to the 
superconducting layers)\cite{Seow96,Doyle96,Morozov98,Morozov99} and of 
the field at which the Josephson plasma resonance 
occurs in the range 24 - 45 GHz.\cite{Sato97,Kosugi97,Hanaguri97,Kosugi98} 
Measurements of the Josephson plasma resonance and of the $c$-axis 
conductivity,\cite{Morozov98,Morozov99}
which have the advantage of directly probing the Josephson critical current and the 
cosine of the average phase difference between layers, show the existence, 
in the heavy-ion irradiated layered superconductor Bi$_{2}$Sr$_{2}$CaCu$_{2}$O$_{8+\delta}$, 
of a regime $B_{l}(T)<B<B_{h}(T)$ in the vortex liquid state in which interlayer  superconducting
phase coherence increases as the magnetic field is increased ({\em i.e.} a ``recoupling''). 
This behavior is qualitatively opposite to that found in the unirradiated material. 
At fields greater than $B_{h}$, interlayer phase coherence again decreases as field is increased. 

The peculiar field dependence of the $c$-axis 
electromagnetic response, as well as that of the reversible --
and even of the irreversible part of the magnetization 
(below $H_{irr}(T)) $\cite{Chikumoto98} have been interpreted in terms 
of the competition between the Josephson coupling energy, which tends to 
align pancake vortices, and the entropy contribution to the free energy 
arising from a random distribution of pancakes over columnar 
defects.\cite{Bulaevskii98} The importance of intralayer pancake 
repulsion is often downplayed, even though it has been shown to be 
important in determining the pancake arrangement,\cite{Wengel98} reversible 
magnetization,\cite{Wahl95,Wengel98} and transport properties.\cite{Wengel98}
It is the purpose of this paper to investigate the importance of 
the contributions of intra- and interlayer pancake vortex interaction 
energy, entropy, and pinning 
energy to the Gibbs free energy $G$, and the impact of each on the phase 
diagram of heavy-ion irradiated Bi$_{2}$Sr$_{2}$CaCu$_{2}$O$_{8+\delta}$. 
To this effect, we have performed 
reversible magnetization as well as AC screening measurements on 
heavy-ion irradiated Bi$_{2}$Sr$_{2}$CaCu$_{2}$O$_{8+\delta}$ with 
widely varying matching fields (section \ref{section:exp}). 
From the analysis presented below (section \ref{section:rev}), it turns out that in the London 
regime, the entropy contribution to the reversible magnetization is 
rather insignificant with respect to the total magnetization, or its 
modification due to vortex pinning on the columnar defects. Only in 
irradiated crystals with sufficiently high $B_{\phi}$ is it essential 
to take a configurational entropy contribution into account when 
describing the magnetization, and then, only at low fields $B \ll 
B_{\phi}$ (section~\ref{section:entropy}). As a consequence, previous estimates of pinning 
energies\cite{vdBeek96II,QiangLi96,Hardy98} determined from the reversible 
magnetization should be revised towards (sometimes significantly) lower values. Next, it is shown how 
information on pancake alignment can be obtained from the reversible 
magnetization (section~\ref{section:coupling}).
The data allow for the extraction of the field $H_{int}$ at which vortex repulsive interactions 
begin to determine the pancake vortex arrangement over the columnar defects. 
This field correlates very well with sharp features in the 
irreversibility line, and, in some cases, with the ``recoupling'' 
observed in the JPR measurements and $c$--axis conductivity (\ref{section:irr}).

\section{Experimental details}
\label{section:exp}

\subsection{Sample preparation}

For this study, we have used Bi$_{2}$Sr$_{2}$CaCu$_{2}$O$_{8+\delta}$ single 
crystals grown at the University of Amsterdam by the travelling solvent floating zone technique.\cite{TWLi94}
A number of crystals were postannealed at 800$^{\circ}$C in air and had $T_{c} = 90.0$ K.
Other crystals were retained as-grown --- this yields samples that 
are lightly overdoped in oxygen, and that have $T_{c} \approx 83$ K. 
Another batch of crystals, grown at the University of Tokyo using the 
same technique,\cite{Motohira} had $T_{c} = 88.8$ K and was used 
for the measurements of the irreversibility line for different matching fields.
After growth and annealing, large crystals were selected for uniformity and absence of 
macrodefects, and subsequently cut using a wire saw, so as to produce small 
squares of typical size $800 \times 800 \times 20$ $\mu{\rm m}^{3}$, suitable
for magnetometry experiments. A number of larger pieces ($1 \times 2$ 
${\rm mm}^{2} \times 20 $ $\mu{\rm m}$) of the Amsterdam crystals were kept for 
reversible magnetization measurements using a commercial superconducting quantum 
interference device (SQUID) magnetometer, as well as ac transmittivity measurements. 
In this manner, a sizeable number of crystals with similar characteristics was obtained.

The samples were irradiated with 5.8 GeV Pb ions at the Grand 
Acc\'{e}l\'{e}rateur National d'Ions Lourds (GANIL) at Caen, France. 
Such an irradiation produces continuous amorphous tracks of radius 
$c_{0} \sim 3.5$ nm which traverse the samples along their entire 
thickness. During the irradiation, the ion beam traverses a 1 $ \mu$m 
thick Ti film placed in front of the sample; the secondary electron emission 
from this film is used to continuously monitor the ion flux during 
irradiation. At the beginning of each run, the ion flux is  calibrated 
using a Faraday cup placed between the Ti-film and the sample. This procedure 
allows one to accurately control the total ion fluence for each sample. 
Moreover, the ion beam is continuously swept across the target using 
asynchronous vertical (3 Hz) and horizontal (1 kHz) ac drive fields so 
as to expose the entire target area ($2 \times 3$ ${\rm 
cm}^{2}$) homogeneously. Different crystals from each source were irradiated 
with widely different ion fluences, in order to produce samples with dose 
equivalent matching fields 0.02 T$ < B_{\phi} < 4$ T. There was a 
slight reduction of the samples' $T_{c}$ after the irradiation. 
Notably, overdoped single crystals had a $T_{c}$ of 81 K and 78.7 K 
after irradiation to doses corresponding to $B_{\phi} = 2$ and 4 T 
respectively. The crystals from the University of Tokyo underwent a 
reduction of $T_{c}$ of up to 4 K (for $B_{\phi} = 4$ T).

\subsection{ac shielding}
\label{section:transmittivity}

The ac transmittivity\cite{Gilchrist93} was measured using the Local Hall Probe 
Magnetometer, in the same way as described in Ref. \onlinecite{vdBeek95II}. 
The sample was placed on top of a miniature InSb Hall sensor; both 
were surrounded by an AC drive coil that could produce fields $h_{ac}$ of up to 
30 G (at frequencies 0.5 Hz $< f <$ 2 kHz). The ac field was always directed parallel 
to the sample $c$-axis. This corresponds to the application to the sample of an 
azimuthal electric field  $E_{\phi}$ with gradient 
$r^{-1} \partial( r E_{\phi} )/\partial r = 2\pi h_{ac}f$ ($r$ is the radial 
coordinate). The ac Hall voltage $V_{ac}$ was measured as function of temperature 
and applied field $H_{a}$. From $V_{ac}$, the first and third harmonic components of the 
transmittivity are determined as $T_{H}(f,T) \equiv 
(V_{ac}(f,T) -V_{ac}(f,T\! \ll \!T_{c}))/(V_{ac}(f,T \! \gg \! T_{c})-V_{ac}(f,T  \! \ll \! 
T_{c}))$ and $T_{H3}\equiv V_{ac}(3f,T)/(V_{ac} (f,T \! \gg \! T_{c}) - 
V_{ac}(f,T \! \ll \!T_{c}))$ respectively. In order to determine the irreversibility field 
$H_{irr}(T,f)$ (or irreversibility temperature $T_{irr}(H_{a},f)$), the transmittivity was 
measured using an ac field of amplitude $h_{ac} = 1$ G and frequency $f = 
7.75$ Hz. $T_{irr}(H_{a},f)$ was taken as the temperature at which the $T_{H3}$ signal first 
becomes distinguishable from the background noise when cooling. Given 
the size of our samples, this corresponds to probing whether the $E(j)$--characteristic deviates from Ohm's law 
at an electric field value of 0.6 nVcm$^{-1}$. In practice, the measurement of $T_{irr}$ is 
often limited by instrumental resolution, which can be estimated by 
comparing the signal at full screening to the noise level; it 
corresponds, for a sample of area $800 \times 800$ $\mu{\rm m}^{2}$, to a circulating shielding 
current density of $4 \times 10^{2}$ ${\rm Am^{-2}}$.
 
\subsection{Reversible magnetization}

Measurements of the reversible magnetization were carried out using a 
commercial SQUID magnetometer at the University of Leiden. Care was taken 
to avoid the numerous artefacts previously observed in such measurements. 
Noteably, we have found that the reported ``clockwise'' 
hysteresis\cite{Pradhan95} is an artefact related to the use of a too restricted length over which the 
sample is drawn through the SQUID gradiometer coils when the 
magnetic moment is small. A scanlength of 6 cm produced 
satisfactory results, with a reproducible reversible magnetic moment at 
fields greater than $H_{irr}(T)$. However, it is well-known that, as a 
consequence of the inhomogeneity of the field produced by the superconducting solenoid, 
measuring the magnetization with such a scanlength at fields close too 
but \em below \rm $H_{irr}$ corresponds to cycling the sample through minor 
magnetic hysteresis loops. The magnetic moment then results from 
the superposition of several ill-controlled shielding current loops 
and can take on a value that is very different from that expected 
from the ordinary full critical state. It thus proved very difficult 
to accurately measure the magnetic moment both above \em and \rm below  
$H_{irr}(T)$; for the present paper, we chose conditions such as to 
obtain the most reliable results above $H_{irr}(T)$, and to rely on the 
ac transmittivity measurements to obtain the screening current below 
$H_{irr}(T)$.

\section{Results and data analysis}

\subsection{Reversible magnetization}
\label{section:rev}
\begin{figure}
	\centerline{\epsfxsize 8.5cm \epsfbox{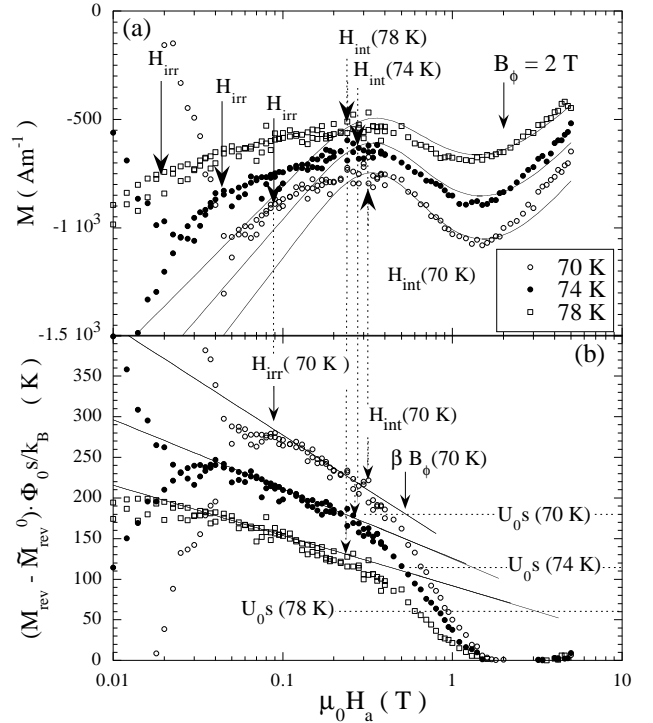}}
	\caption{(a) The reversible magnetization of a heavy-ion irradiated optimally 
	doped Bi$_{2}$Sr$_{2}$CaCu$_{2}$O$_{8+\delta}$ single crystal
	($B_{\phi} = 2$ T;	$T_{c} = 88.8$ K), for temperatures T = 70, 74, and 78 K. 
	The drawn lines represent 
	fits to Eq.~(\protect\ref{eq:Buzdin}) with parameter values 
	$\lambda_{L}(0) = 240$ nm ($\lambda(0) = 180$ nm), $\partial B_{c2}/\partial T= -1.05$ T/K,
	and $U_{0}s \sim 1.5 \tilde{\varepsilon}_{0}s \approx
	1300 (1-T/T_{c})$ K.  Arrows indicate the field below which the ac 
	transmittivity has a non-zero third harmonic ($H_{irr}$) and the 
	interaction field $H_{int}$	above which vortex interactions  influence the vortex 
	arrangement;
	(b) ($\Phi_{0} s / k_{B} = 0.23$ Km/A) times the difference between the measured magnetization and the 
	theoretical magnetization in the absence of columnar defect pinning, $\tilde{M}_{rev}^{0}$ 
	(corrected for the reduction in superfluid density expressed by $\tilde{\lambda}^{-2}(T) = 
    (1 - 2\pi c_{0}^{2}n_{d}) \lambda^{-2})$. In the limit $B \ll 
    B_{\phi}$, the plotted data correspond exactly to the sum of 
	pinning energy and entropy per pancake vortex. Horizontal dotted lines correspond to the \em 
	bare \rm pinning energy $U_{0}s = U_{0}(0)s(1-T/T_{c})^{2}$, with 
	$\tilde{U}_{0}(0)s = 4000$ K. This corresponds to the parameter value obtained in 
	Ref.~\protect\onlinecite{Drost99}, corrected for the change 
	$\lambda \rightarrow \tilde{\lambda}$. The drawn lines show the 
	logarithmic extrapolation of the low field data which permit the 
	determination of the length $L_{a}$ over which pancakes 
	belonging the same vortex line are aligned on the same column, and 
	the factor $\beta$ describing the fraction of available columns 
	(see section \protect\ref{section:coupling}). 
	For $T = 70$ K, the low field data obey $U_{0}s + 
	T \ln(\beta B_{\phi}/B)$, \em i.e. \rm the typical stacklength 
	$L_{a}$ is close to the interlayer spacing; for larger temperatures the prefactor 
	to the logarithm is, unexpectedly, somewhat less. The interaction field $H_{int}$ of 
	Fig.~\protect\ref{fig:Mrev-T}(a) 
	is determined as that where the data first deviate from a logarithmic $H_{a}$--dependence.}
	\label{fig:Mrev-T}
	\end{figure}

Figure \ref{fig:Mrev-T} shows the reversible magnetization for an 
optimally doped Bi$_{2}$Sr$_{2}$CaCu$_{2}$O$_{8+\delta}$ single 
crystal ($T_{c} = 88.8$ K),\cite{VB5} 
irradiated with $10^{11}$ Pb ions ${\rm cm^{-2}}$, which 
corresponds to a matching field $B_{\phi} = 2$ T. This value of 
$B_{\phi}$ is comparable to that used in previous 
studies,\cite{Chikumoto98,Wahl95,vdBeek96II,QiangLi96,Drost98} and the reversible 
magnetization shows all the features discussed there. 
Recapitulating, at low fields $H_{irr}< 
H_{a} \ll B_{\phi}$, the absolute value of the magnetization
$|M_{rev}|$ decreases more or less proportionally to $\ln(1/H_{a})$, in

\noindent   agreement 
with the London model which has $M_{rev} \approx - (\varepsilon_{0}/2\Phi_{0}) \ln(\eta 
B_{c2}/eB) \equiv M_{rev}^{0}$. Here $\varepsilon_{0} = 
\Phi_{0}^{2}/4\pi\mu_{0}\lambda^{2}(T)$, 
$\lambda(T)$ is the penetration depth, $B_{c2}$ is the upper 
critical field, $\mu_{0} = 4\pi \times 10^{-7}{\rm Hm^{-1}}$, $\eta 
\sim 1 $ and

\begin{figure}
\centerline{\epsfxsize 8.5cm \epsfbox{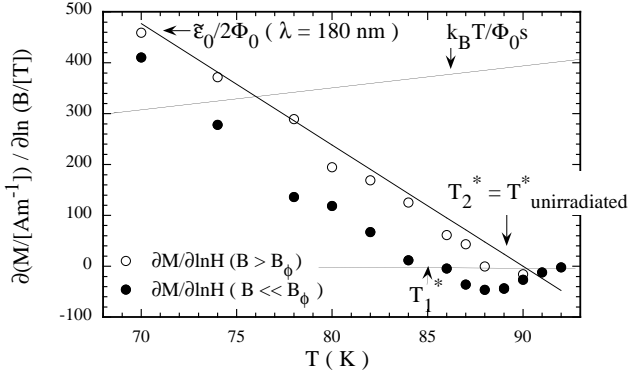}}
\caption{Logarithmic slopes $\partial M_{rev} / \partial \ln B$ 
for the same crystal as in Fig.~\protect\ref{fig:Mrev-T}, determined 
in the limits $B > B_{\phi} = 2$ T and $B \ll B_{\phi}$ respectively. 
The drawn lines indicate the magnitude of the theoretical entropy 
contribution in the 2D limit, $k_{B}T/\Phi_{0}s$, and the prediction 
of the London model, $M^{0}_{rev}=\tilde{\varepsilon}_{0}/2\Phi_{0}$.}
\label{fig:dM/dlnH}
\end{figure}

\noindent $e = 2.718\ldots$. Throughout, the small magnitude 
of $M_{rev}$ allows us to take $B \approx \mu_{0}H_{a}$. The
logarithmic decrease of $|M_{rev}|$ terminates at the 
field labelled $H_{int}$, whence $|M_{rev}|$ increases. When $B 
\gtrsim B_{\phi}$, $|M_{rev}|$ decreases once again proportionally to 
$\ln(1/B)$, although the slope $\partial M_{rev}/\partial \ln B$ 
is markedly higher than the slope at low fields.
The logarithmic slopes in either limit are illustrated in Fig.~\ref{fig:dM/dlnH}, which shows  
$\partial M_{rev}/\partial \ln B$ as 
function of temperature for the same crystal: for fields $B \ll B_{\phi}$
the logarithmic slope lies markedly below that measured for $B > B_{\phi}$; 
also, it goes to zero at a temperature $T^{*}_{1}$ that is lower 
than the temperature $T^{*}_{2}$ at which $\partial M_{rev}(B > 
B_{\phi})/\partial B$ goes to zero. Both $T^{*}_{2}$ and $\partial M_{rev}(B > 
B_{\phi})/\partial B$ are very similar to the corresponding values measured in the 
unirradiated crystal.\cite{vdBeek96II} In both cases, unirradiated 
crystals and irradiated crystals in the high field limit, one can use the 
London model\cite{Clem,Koshelev94II} to extract the value of the penetration depth
from $\partial ^{2} M_{rev}(B > B_{\phi})/\partial T \partial \ln 
B = - \varepsilon_{0}(0)/2\Phi_{0}T_{c}$, an  exercise which yields 
$\lambda(0) = 180 \pm 30$ nm. The corresponding value of the London penetration depth 
$\lambda_{L}(0) \approx 1.35 \lambda(0) \approx 240$ nm is in good agreement with 
literature values for optimally doped 
Bi$_{2}$Sr$_{2}$CaCu$_{2}$O$_{8+\delta}$.\cite{TWLi96} The error (of less than 20 \%)\cite{Kogan93}
arises because of two reasons. First, the presence of the normal cores 
of the columnar defects decreases the supercurrent density around each 
vortex and the strength of intervortex repulsion. This effect can be 
taken into account by a ``renormalized'' value $\tilde{\lambda}^{-2}(T) = 
(1 - 2\pi c_{0}^{2}n_{d}) \lambda^{-2}$ and corresponding 
$\tilde{\varepsilon}_{0} = \Phi_{0}^{2}/4\pi\mu_{0}\tilde{\lambda}^{2}$ 
and $\tilde{M}_{rev}^{0} = \tilde{\varepsilon}_{0}/2\Phi_{0} \ln( \eta 
B_{c2}/ e B)$.\cite{Wahl95} 
More important, the presence of thermal fluctuations was predicted to lead 
to a substantial entropy contribution to the free energy, reflected by an 
extra term:\cite{Bulaevskii92} 

\begin{equation}
	\frac{\partial ^{2} M_{rev}}{\partial T \partial \ln
B} = \frac{\partial ^{2} M_{rev}^{0}}{\partial T \partial \ln
B} - \frac{k_{B}}{\Phi_{0}s}. 
\end{equation}

\noindent Here, $s \approx 1.5$ nm is half the $c$--axis parameter
 of the Bi$_{2}$Sr$_{2}$CaCu$_{2}$O$_{8+\delta}$ material
 ({\em i.e.} the distance between CuO$_{2}$--bilayers). From the value 
$k_{B}/\Phi_{0}s = 4.4$ ${\rm Am^{-1}K^{-1}}$, we see that the (constant) 
contribution of the entropy term\cite{Bulaevskii92} 
  can amount to 20 \% of the temperature derivative 
of $\partial M_{rev}/\partial \ln B$, at most.

The nonmonotonic field dependence of the reversible magnetization has a 
straightforward explanation in terms of the lowering of the vortex 
free energy by pinning onto the columnar 
defects.\cite{vdBeek96II,Drost98} The magnetization $M_{rev} = - 
\partial G / \partial B = - \Phi_{0}^{-1} \partial G / \partial 
n_{v} \equiv - \Phi_{0}^{-1} \mu$, with $n_{v}$ the vortex density, corresponds to the vortex chemical 
potential $\mu$ and therefore directly measures the energy needed to add 
a vortex to the system when the field is increased. At low fields many columnar defects 
are available, so that each and every new vortex can gain a maximum energy per unit length
$U_{0}$ by becoming trapped in the defect potential; correspondingly 
$M_{rev} \sim \tilde{M}_{rev}^{0} + U_{0}/\Phi_{0}$.  
The pinning of a vortex on an insulating columnar defect 
entails the redistribution of the supercurrent towards the vortex 
periphery, lower maximum current on the vortex core boundary, and a 
smaller intervortex repulsion. More vortices can therefore enter the 
sample at fixed magnetic field, hence the absolute value of the magnetization is smaller. 
However, it is still expected that $M_{rev} \propto \ln(1/B)$ in 
this regime.\cite{Wahl95,vdBeek96II,Bulaevskii96II} 
At high fields $\mu_{0}H_{a} > B_{\phi}$ nearly all columns 
are occupied, new vortices cannot be 
trapped on a defect, hence $M_{rev} \approx \tilde{M}_{rev}^{0}$, 
close to the magnetization of an unirradiated 
sample in the London regime.\cite{vdBeek96II} In the intermediate field 
region, $\mu_{0}H_{int} < \mu_{0}H_{a} < B_{\phi}$ in 
Fig.~\ref{fig:Mrev-T}(a), new vortices can 
become either trapped or free. The proportion of trapped vortices 
decreases as field increases, leading to the increase of $|M_{rev}|$. 

\subsubsection{Role of intervortex repulsion}

The magnetization in the intermediate regime has been estimated in 
Ref.~\onlinecite{Wahl95} by considering that the gain in pinning energy 
for occupation of a particular column should  be larger than the 
loss in energy due to intervortex repulsion when a vortex is moved 
from its lattice site to the columnar track; assuming a 
Poisson-distribution of distances between tracks, this gives

\begin{equation}
	M_{rev} \approx \tilde{M}_{rev}^{0}	
	   - U_{0}\left[ 1-\left( 1+ \frac{U_{0}}{\tilde{\varepsilon}_{0}} \frac{B_{\phi}}{B}\right) {\rm 
	   e}^{-(U_{0}/\tilde{\varepsilon}_{0})(B_{\phi}/B)} \right].
 \label{eq:Buzdin}
\end{equation}

\noindent In spite of its simplicity, it is possible to obtain
acceptable fits to the experimental data at $H_{a} \gtrsim H_{int}$ using this expression  
(see Fig.~\ref{fig:Mrev-T}(a) and Ref.~\onlinecite{Hardy98}). In the 
Figure we have used $\partial B_{c2} / \partial T = -1.05 $ T/K, \cite{Drost99} 
and $\lambda(0) = 180$ nm as determined above.  
However, for a faithful fit to the peak structure of $|M_{rev}|$ 
one should assume $U_{0}/\tilde{\varepsilon}_{0}$ to be in excess 
of unity; for example, at 70 K, $U_{0}/\tilde{\varepsilon}_{0} = 1.5 $ in Fig.~\ref{fig:Mrev-T}. 
Such large values of $U_{0}$ were also found in 
Refs.~\onlinecite{QiangLi96} and \onlinecite{Hardy98}; they are incompatible with 
the theoretical expectation for electromagnetic pinning of vortices,
\cite{Hardy98,Mrktchyan71,Buzdin94} $U_{0} \approx 
\tilde{\varepsilon}_{0} \ln ( c_{0}/\sqrt{2}\xi)$, since they would imply that 
the column radius $c_{0}$ exceeds the 
coherence length $\xi = 2.2 (1-T/T_{c})^{-1/2}$ nm by a factor 5 -- 6. In 
Ref.~\onlinecite{Drost99}, it was found that pinning of the vortex core dominates over 
electromagnetic pinning in heavy-ion irradiated 
Bi$_{2}$Sr$_{2}$CaCu$_{2}$O$_{8+\delta}$ at all temperatures. 
Then, $U_{0} \approx \varepsilon_{0}  ( c_{0}/2 \xi )^{2}$, 
yielding $c_{0} \approx 2.3 \xi(0)$. This value is still somewhat high, 
which shows that the actual pinning mechanism is more complicated 
than simple core pinning. However, note that even so the 
magnetization data at fields $H_{a} \ll H_{int}$ lie much above 
the theoretical curve. 

We thus find that a description that only takes the decrease of the 
internal energy due to the localization of the vortices on the 
columnar defects into account is insufficient to describe the large 
decrease of the low-field magnetization at large $B_{\phi}$. The fact that 
the low-field logarithmic slope $\partial M / \partial \ln B$ is much 
smaller than that predicted by Eq.~(\ref{eq:Buzdin}) indicates the presence 
of an extra contribution to the free energy that has a
logarithmic dependence $\sim \ln(1/B)$. 

\subsubsection{Role of entropy}
\label{section:entropy}

In this section we examine the importance of a possible entropy contribution 
$-TS$ to the free energy. Such a contribution may arise from
vortex positional\cite{Koshelev94II,Bulaevskii92,Bulaevskii96II} and order parameter 
amplitude fluctuations.\cite{Koshelev94II,Ikeda95,Tesanovic92} The 
entropy contribution to the free energy is believed to be important in layered 
superconductors because of the ``crossing point'' behavior observed 
in the reversible magnetization at $T_{1}^{*}$ and 
$T_{2}^{*}$.\cite{Chikumoto98,vdBeek96II,Kes91,QiangLi93} 
Supposing that a description in terms of the London model is 
appropriate, and hence, that vortex positional fluctuations give the dominant contribution to $S$, 
it was derived in Ref.~\onlinecite{Bulaevskii92} that the crossing point arises
in the unirradiated material when the 
field dependence of the two terms contributing to $M_{rev}$,

\begin{equation}
M_{rev}^{un}   =   M_{rev}^{0}	 
                   + \frac{k_{B}T}{\Phi_{0}s} \ln \left(\frac{B_{0}}{B} \right)	
\label{eq:Bulaevskii} 
\end{equation}

\noindent with derivative

\begin{equation}
\frac{\partial M_{rev}^{un}}{\partial \ln B}   =     
                     \frac{\varepsilon_{0}}{2\Phi_{0}} - \frac{k_{B}T}{\Phi_{0}s}
\end{equation}

\noindent cancel at $T^{*} = T_{c}/(1 + 
2k_{B}T_{c}/\varepsilon_{0}(0)s)$. Here $B_{0}$ is a field determined by 
the elemental phase area;\cite{Koshelev94II,Bulaevskii92} the 
fraction $B_{0}/B$ corresponds to the total number of possible states 
(``configurations'') per vortex. In two dimensions (2D), $B_{0} \approx 
B_{c2}$.\cite{Koshelev94II} In heavy-ion irradiated layered superconductors, Eq.~(\ref{eq:Bulaevskii}) 
should be modified: the full result can be found in Ref.~\onlinecite{Bulaevskii96II},
but the limiting forms for $M_{rev}$ are:

\begin{eqnarray}
M_{rev} & \approx &     \tilde{M}_{rev}^{0} + \frac{U_{0}}{\Phi_{0}}  + 
	                    \frac{k_{B}T}{\Phi_{0}s}\ln \left(\frac{B_{\phi}}{B}\right)
\hspace{7mm}( B \ll B_{\phi} )
\label{eq:Bulaevskiic1} \\
M_{rev} & \approx  & \tilde{M}_{rev}^{0}	 
                   + \frac{k_{B}T}{\Phi_{0}s} \ln \left(\frac{B_{0}}{B} \right)	
\hspace{3mm}(B \gg B_{\phi} ; U_{0} \rightarrow 0 ) .
\label{eq:Bulaevskiic2}
\end{eqnarray}

\noindent These are again valid in the 2D limit where the positions of pancake vortices 
in neighboring layers are uncorrelated. The logarithmic field dependence of the second term of 
Eq.~(\ref{eq:Bulaevskiic1}) now depends on the number ($\sim B_{\phi} / 
B$) of columnar defect sites available to each pancake vortex.

In order to estimate the importance of the entropy 
contribution to the experimentally measured magnetization, we return to 
the temperature dependence of the logarithmic slopes $\partial 
M_{rev}/\partial \ln B$ 
in Fig.~\ref{fig:dM/dlnH}. Note that, according 
to Eqs.~(\ref{eq:Bulaevskii}--\ref{eq:Bulaevskiic2}), $\partial 
M_{rev}/\partial \ln B$ is equal to 
the difference between two large terms: 

\begin{equation}
\frac{\partial M_{rev}}{\partial \ln B} =   
\frac{\partial \tilde{M}_{rev}^{0}}{\partial \ln B} - \frac{k_{B}T}{\Phi_{0}s}.
\end{equation}

\noindent However, these terms have very 
different temperature derivatives, equal to $-\tilde{\varepsilon}_{0}(0)/2\Phi_{0}T_{c}$
 and $- k_{B}/\Phi_{0}s$ respectively. The ratio between the temperature
derivatives is equal to $2 Gi^{1/2}$, where $Gi$ is the 2D Ginzburg 
number that determines the  regime of reduced temperature $t(T) = (T_{c}-T)/T_{c}$ over which the 
entropy contribution to the free energy is dominant. For optimally doped
Bi$_{2}$Sr$_{2}$CaCu$_{2}$O$_{8+\delta}$, $2k_{B}T_{c}/\varepsilon_{0}(0) s 
\approx 0.19$, {\em i.e.} $Gi \approx 0.01$; for our overdoped 
material, $Gi \approx 0.003$. The value 0.19 means that the error in the 
$\tilde{\lambda}(0)$--value deduced from the reversible magnetization, 
arising from an entropy contribution to the free energy,
cannot exceed 19\% (11\% for the overdoped crystals). 
Inserting the experimental $\tilde{\lambda}$--value back into $\tilde{M}_{rev}^{0}$, 
and comparing the result to the full measured magnetization, we 
find that for $B \gg B_{\phi}$ as well as for unirradiated crystals,\cite{Kes91} the 
entropy contribution to $\partial M_{rev}/\partial \ln B$ does not exceed 0.1
$k_{B}T/\Phi_{0}s$ (see Fig.~\ref{fig:dM/dlnH}). This excludes 
an interpretation of the crossing point in 
terms of quasi--2D vortex positional fluctuations only: for example, 
Eq.~(\ref{eq:Bulaevskii}) would lead to $T^{*} = 75$ K.  More 
generally, vortex translational degrees of freedom do not lead to a major 
modification of the magnetization of unirradiated crystals (or 
irradiated crystals at fields above $B_{\phi}$) in the London regime.
The entropy contribution to the free energy only starts to be important in the 
critical fluctuation regime.

With respect to this, we note that the analysis of 
Ref.~\onlinecite{Kes91} has demonstrated that the crossing point in unirradiated 
Bi$_{2}$Sr$_{2}$CaCu$_{2}$O$_{8+\delta}$ crystals indeed lies in the 
regime of (2D) critical fluctuations. An adequate description, valid 
for  $B > \frac{1}{3} t (\partial B_{c2} / \partial 
t)_{T=T_{c}}$, was obtained using in Ref.~\onlinecite{Tesanovic92}
by expanding the order parameter in terms of lowest Landau level 
(LLL)--eigenfunctions. The expression for $M_{rev}$ derived in 
Ref.~\onlinecite{Tesanovic92} (which also predicts $\partial M_{rev}/\partial 
\ln B \rightarrow \varepsilon_{0}/2\Phi_{0}$ for $t \gg t(T^{*})$) well 
describes the experimental data for both unirradiated 
Bi$_{2}$Sr$_{2}$CaCu$_{2}$O$_{8+\delta}$ \cite{Tesanovic92} 
and irradiated Bi$_{2}$Sr$_{2}$CaCu$_{2}$O$_{8+\delta}$ above the matching 
field.\cite{vdBeek96II} In Fig.~\ref{fig:dM/dlnH}, the $B \gg B_{\phi}$--data 
lie outside the London limit for $T \gtrsim 80 K$, and 
within the LLL--regime for $T \gtrsim 84$ K. An interpretation of the 
field--independent magnetization at $T_{2}^{*}$ in terms of vortex 
positional fluctuations (order parameter phase fluctuations) only 
is thus inappropriate.

The situation is totally different for $B \ll B_{\phi}$,
at which one has a significantly larger 
value $TS \gtrsim 0.3 k_{B}T/\Phi_{0}s$. This means that 
the introduction of many extra possible (columnar defect) sites 
within a vortex lattice unit cell leads to a large 
(configurational) entropy contribution. Moreover,  
a description in terms of the London model will turn out to be adequate in this field regime.

\subsubsection{Interlayer coupling; dependence on matching field}
\label{section:coupling}

The small entropy contribution to the reversible magnetization for 
$B \gtrsim B_{\phi}$ suggests that even in the vortex liquid, significant 
positional correlations remain between vortex pancakes in adjacent layers.
In that case, $S$ is reduced with respect to its value in
the 2D limit: the alignment of pancakes in different layers means 
that the relevant entities are stacks (of length $L_{a}$) comprising several pancakes. 
The entropy contribution in the regime $B \ll B_{\phi}$ where all 
pancakes are trapped by a column  can be estimated as 
$TS = k_{B}T\ln W =  k_{B}T\ln N_{c}^{L_{s}/L_{a}}$, with $N_{c} 
\equiv \beta B_{\phi}/B$ the number of columns available to each stack 
of pancakes, $L_{s}$ the length of the sample in the field direction, and $L_{a}$ the length 
$\parallel c$ over which the pancakes are aligned on the same columnar 
track. The number $1-\beta$ is the fraction of columns that is
inaccessible due to intervortex repulsion. In the quasi--2D limit, $L_{a} = s$, 
but in general $L_{a}$ can be estimated from the balance between elastic 
energy and thermal energy: $\tilde{\varepsilon}_{1} \Phi_{0}/ B_{\phi} L_{a} \sim k_{B}T$ 
($\tilde{\varepsilon}_{1} \approx \tilde{\varepsilon}_{0}/\gamma^{2}$ is the line 
tension and $\gamma$ the anisotropy parameter).
Hence the entropy contribution per unit length $TS = k_{B}T\ln N_{c}^{ 
L_{s}n_{d} k_{B}T/ \tilde{\varepsilon}_{1}}$ and the magnetization\cite{Blatter}

\begin{eqnarray}
M_{rev} & =  &  M_{rev}^{0} + \frac{U_{0}}{\Phi_{0}} + 
	                \frac{k_{B}T}{\Phi_{0}s}    
	                \left(\frac{k_{B}T}{\tilde{\varepsilon}_{0}s} \right)  
                 	\frac{B_{\phi}}{B_{cr}} \ln \left( \frac{\beta B_{\phi}}{B}\right) 
					\label{eq:correlated}\\
        & =   & M_{rev}^{0} + \frac{U_{0}}{\Phi_{0}} + 
                	\frac{k_{B}T}{\Phi_{0}s} \frac{s}{L_{a}}
					\ln\left( \frac{\beta B_{\phi}}{B}\right) \label{eq:length} \\
		&	&	\hspace{3.5cm}	( B_{\phi}< B_{cr} \tilde{\varepsilon}_{0}s/k_{B}T )
				\nonumber	. 
\end{eqnarray}

\noindent The crossover field $B_{cr} \equiv 
\Phi_{0}/(\gamma s)^{2}$ delimits the (low field) regime where tilt 
deformations of the vortex lattice with wavevector $\sim 4\pi (B/\Phi_{0})^{1/2}$ 
are more difficult to excite than shear deformations with similar 
wavevector, from the high field regime where the inverse is true. The 
contribution $TS$ adds a supplementary logarithmic dependence on $B$ at low fields, in 
agreement with the difference between the experimental $\partial M_{rev}/\partial 
\ln B$ observed at $B \ll B_{\phi}$ and $B > B_{\phi}$ respectively 
(see fig. \ref{fig:Mrev-T}(a)).

In Ref.~\onlinecite{Chikumoto98}, it was claimed that the maximum in 
$|M_{rev}|$ can be nearly entirely explained by the reduction in $TS$, caused by pancake 
alignment (``recoupling'') and the concommitant decrease in $s/L_{a}$. 
This claim can be tested by using Eq.~(\ref{eq:length}), which permits a direct estimate 
of the correlation length $L_{a}$. Starting from the values of  
$\partial M_{rev}/\partial \ln B$ determined above, one would have

\begin{figure}
	\centerline{\epsfxsize 8cm \epsfbox{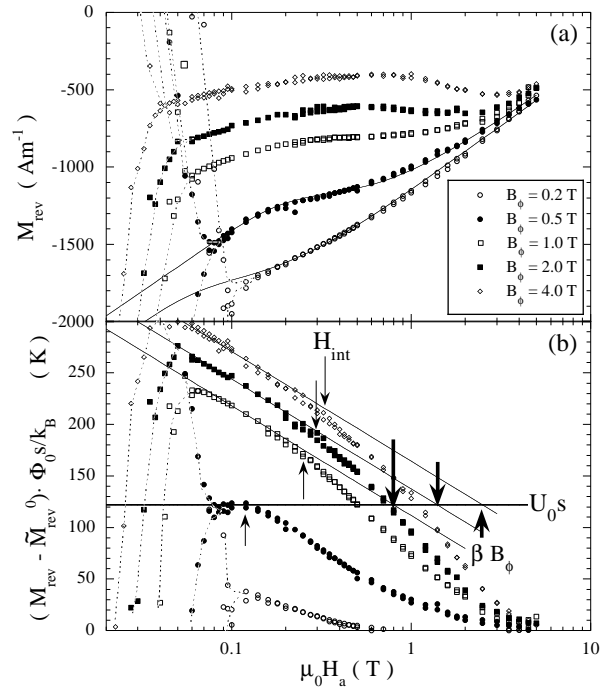}}
	\caption{(a) Reversible magnetization of lightly overdoped Bi$_{2}$Sr$_{2}$CaCu$_{2}$O$_{8+\delta}$
	at $T/T_{c} = 0.89$ , for matching fields 0.2 T $< B_{\phi} <$ 4 T  .
	The critical temperature $T_{c} \approx 83.1$ K, except for $B_{\phi} = 
	2$ T ($T_{c} = 81$ K )
	and $B_{\phi} = 4$ T ($T_{c}$ = 78.7 K). The drawn lines through 
	the data points for $B_{\phi} = 0.2$ and 0.5 T represent 
	fits to Eq.~(\protect\ref{eq:Buzdin}) with parameter values 
	$\lambda(0) = 140$ nm and $U_{0}s = 120$ K; the dotted lines are guides to 
	the eye, representing the magnetization in the irreversible regime. 
	(b) $\Phi_{0} s / k_{B}$ times the difference between the measured magnetization and the 
	theoretical magnetization in the absence of columnar defect pinning, $\tilde{M}_{rev}^{0}$, 
	(corrected for the reduction in superfluid density expressed by $\tilde{\lambda}^{-2}(T) = 
    (1 - 2\pi c_{0}^{2}n_{d}) \lambda^{-2})$. In the limit $B \ll 
    B_{\phi}$, the plotted data correspond exactly to the sum of 
	pinning energy and entropy per pancake vortex. The drawn lines show the \em 
	bare \rm pinning energy per pancake vortex $U_{0}s$ 
	(corresponding to the low field limit of the $B_{\phi} = 0.5$ T --data), and, for 
	$B_{\phi}\geq 1.0$ T, $U_{0}s$ plus the 
	configurational entropy contribution $ TS \approx 
	(k_{B}T/\Phi_{0}s)\ln(\beta B_{\phi}/B)$
	arising from the possibility of individual pancake 
	vortices to occupy different column sites. The fraction $\beta \approx 1/2$ 
	can be	extracted from the intercept with the 
	horizontal $U_{0}s$-line, and is indicated by the 
	bold arrows. The fields $H_{int}$ where intervortex interactions 
	begin to influence the magnetization can be determined as the field 
	of first deviation from a logarithmic field dependence and are 
	indicated using the thin arrows. Dotted lines are guides to the eye, 
	indicating data in the irreversible regime.}
	\label{fig:Mrev-Bfi}
	\end{figure}

\noindent $L_{a} \lesssim  s/0.3 \approx 3 s$ for $\mu_{0}H_{a} \ll B_{\phi} = 2$ T, 
and $L_{a} \gtrsim 10 s$ for $\mu_{0}H_{a} \gtrsim B_{\phi}$, similar
to the value $L_{a} \approx 15 s$ recently obtained by Morozov {\em et 
al.},\cite{Morozov99} and in qualitative agreement with the assumption 
of Ref.~\onlinecite{Chikumoto98}.	

A more stringent approach is to 
consider the dependence of $M_{rev}$ on matching field, {\em i.e.} 
the dependence on the number of available 
states. Figure~\ref{fig:Mrev-Bfi}(a) shows the magnetization of a 
series of lightly overdoped Bi$_{2}$Sr$_{2}$CaCu$_{2}$O$_{8+\delta}$ 
crystals ($T_{c} \approx 83$ K, $\lambda(0) \approx 140$ nm), irradiated 
to different ion fluences corresponding to matching fields ranging between 0.2 T and 4.0 T. 
The Figure shows that there is a large dependence of the 
low--field magnetization on the density of columns, which
irrevocably indicates the importance of the configurational entropy contribution 
to $M_{rev}$ in that regime. Namely, if $S$ were zero, the low--field limits of $M_{rev}$ would 
have been equal to $M_{rev}^{0}+U_{0}/\Phi_{0}$, independent of 
$B_{\phi}$ (see. Eqs.~(\ref{eq:Buzdin},\ref{eq:Bulaevskiic1})).
Furthermore, the data at $B_{\phi} \geq 1$ T confirm the extra logarithmic 
field dependence (\ref{eq:correlated}) for $B \ll B_{\phi}$.
For the crystal with $B_{\phi} = 0.5$ T, the low-- and high field limits 
of $\partial M_{rev}/\partial \ln B$ are nearly equal, implying that 
at this temperature, the entropy contribution to $\partial M_{rev}/ \partial \ln B$ lies within 
the error bar of $0.1k_{B}T/\Phi_{0}s$ found in Section~\ref{section:entropy} (see also Ref.~\onlinecite{Drost98}, 
Fig. 3). Hence, for this matching field and temperature, it follows 
from Eq.~(\ref{eq:length}) that pancakes are aligned into 
stacks of length $L_{a} \gtrsim 10 s$ over the entire considered field 
range. It means that for low matching fields it is allowed to ignore 
the entropy contribution. One can then confidently determine 
$U_{0}s$ from the difference between the measured 
magnetization and the low field limit of $\tilde{M}_{rev}^{0}$.\cite{Drost99,0.2T}
Inserting the obtained value, $U_{0}s 
\approx 120$ K (for $\tilde{\varepsilon}_{0}s \approx 160$ K at 74 K) back into 
Eq.~(\ref{eq:Buzdin}) yields good fits to the data for 
$B_{\phi} = 0.2$ and 0.5 T, as shown by the drawn lines in Fig.~\ref{fig:Mrev-Bfi}. 
Such fits are not possible at higher $B_{\phi}$ due to the 
increasing importance of the entropy contribution, which, even at 
$B_{\phi} = 1$ T already accounts for half the change of the low--field reversible 
magnetization with respect to that of an unirradiated crystal. The 
determination of a ``pinning energy'' from the difference between the low 
field data and the extrapolation to low fields of the data for $B \gg B_{\phi}$ 
can then introduce an error of more than a factor 2.

The parameters $L_{a}$ and $\beta$ can, in principle, be directly
evaluated by subtracting $\tilde{M}_{rev}^{0}$ from the measured magnetization. 
This procedure yields a quantity that at low fields is rigorously 
equal to the sum of the pinning energy per pancake vortex 
and temperature times entropy (Fig. \ref{fig:Mrev-Bfi}(b)). 
The low--field limit of the $B_{\phi} = 0.5 $ T  data corresponds to the 
bare pinning energy $U_{0}s$ --- this value is indicated by the 
horizontal line in Fig.~\ref{fig:Mrev-Bfi}(b). 
The results for the crystals with $B_{\phi} \geq 1$ T exceed $U_{0}s$ 
at $B \ll B_{\phi}$ because of the entropy contribution to the free energy. 
From the logarithmic field derivative and the extrapolation to $U_{0}s$
we find that, at $T/T_{c} = 0.89$, $L_{a}(B \ll B_{\phi}) \approx  s$ 
and $\beta \approx \frac{1}{2}$, independent of $B_{\phi}$. Thus, the magnetization at low fields and 
$B_{\phi} \geq 1$ T is well described by 

\begin{equation}
	M_{rev} = \tilde{M}_{rev}^{0} + \frac{U_{0}}{\Phi_{0}} + 
	\frac{k_{B}T}{\Phi_{0}s} \ln \left( \frac{B_{\phi}}{2 B} \right)
	\end{equation}
	
\noindent as illustrated by the drawn lines through data in Fig.~\ref{fig:Mrev-Bfi}(b)).
The first deviation of the experimental data from these lines as field is increased,
at the ``interaction field'' $H_{int}$, indicates that both configurational entropy and 
the total pinning energy start to decrease due to the effect of 
vortex interactions. Entropy decreases because of the 
limitation of accessible column sites, which is the consequence of intraplane 
pancake repulsion; it can be expressed by a diminishing $\beta$; the translational 
invariance of the system along the column 
direction and the attractive interaction between pancakes in 
adjacent layers will then lead to pancake vortex alignment and the 
supplementary decrease of $S$ through the increase of 
$L_{a}$. The \em total \rm pinning energy $U_{p}$ can only go down 
appreciably if free (or ``interstitial'') vortices, not trapped on a columnar track, start 
to appear, again as a result of intraplane pancake repulsion which 
prohibits certain vortices from finding a favorable column site. From 
Fig.~\ref{fig:Mrev-Bfi}, it is seen that near $H_{int}$, the  
pinning energy (obtained from the $B_{\phi} = 0.5$ T data) and the 
entropy contribution to the free energy are comparable in magnitude 
when $B_{\phi} \gtrsim 1$ T; hence \em both \rm contributions should 
always be taken into account when describing the shape of the 
$M_{rev}(B)$--curve.

We have attempted to carry out the same analysis for constant 
matching fields $B_{\phi} = 0.5$, 1, and 2 T, and different $T$ (see Fig.~\ref{fig:Mrev-T}(b) for 
$B_{\phi} = 2$ T). In this case, the problem arises that the 
bare pinning energy cannot easily be established. It was therefore 
chosen to take the result of Ref.~\onlinecite{Drost99} (where $B_{\phi} = 0.5$ T), 
$U_{0}(T) = U_{0}(0)(1-T/T_{c})^{2}$ for core pinning, and to correct 
the value $U_{0}(0) = 4300$ K obtained there for 
the larger density of columns in those cases where $B_{\phi}> 0.5$ T. 
The difference between the low--field data and these values, represented by 
dashed lines in the Fig.~\ref{fig:Mrev-T}(b), again corresponds to temperature 
times the configurational entropy. For $B_{\phi} = 1$ and 2 T, it is found that 
$TS(B = \beta B_{\phi}/e) \approx T$, indicating that $L_{a} \approx s$. 
The logarithmic field dependence at $T = 74$ and 78 K, as well as the magnitude of the logarithmic 
slopes also agree with $L_{a}/s \approx 1 - 2 $. However, since the 
logarithmic slopes should, in the representation of Fig.~\ref{fig:Mrev-T}(b)), be equal 
to $T$, the predicted temperature dependence is not well obeyed. In 
fact, for the data at $T = 70$ K it is difficult to identify any 
substantial field range over which $TS \propto \ln(1/B)$. This is in
contrast to data for $B_{\phi} = 0.5$ T where a low--field logarithmic 
dependence can be identified at all investigated temperatures.\cite{Drost98,Drost99} 
Hence, accurate values for $H_{int}$ can be determined only for 
small $B_{\phi}$ (see Fig.~\ref{fig:Hint} for $B_{\phi} = 
0.5 $ T). At low $B_{\phi}$, the entropy contribution to the free 
energy is unimportant. The interaction field then delimits 
the low field regime where vortices are pinned independently on the columnar 
tracks, from the high field regime where intervortex repulsion 
entails the presence of unpinned (``free'' or ``interstitial'') vortices.  
Results for larger matching field  can only be obtained in a restricted 
temperature range $T_{irr} < T < T_{c}(1-Gi)$ 
(see Fig.~\ref{fig:phasediag} for $B_{\phi} = 1$ T). Here, the 
interaction field separates the regime of individual vortex pinning 
from high fields at which the intervortex repulsion determines the 
optimum pinned configuration but does not necessarily lead to the 
presence of ``free'' vortices.

We note that, in the analysis presented 
in Fig.~\ref{fig:Mrev-T}(b), the assumption of $U_{0}s 
\approx \varepsilon_{0}(0)s(1-T/T_{c})$ for electromagnetic

\begin{figure}
	\centerline{\epsfxsize 8cm \epsfbox{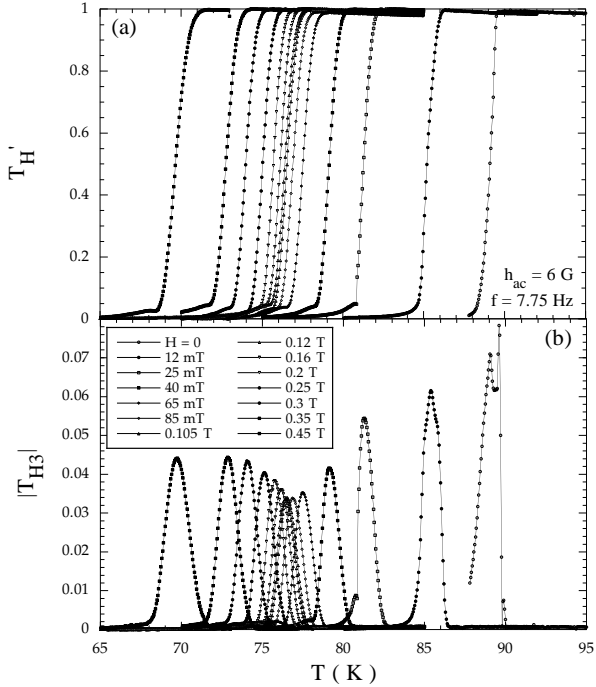}}
	\caption{In-phase fundamental transmittivity $T_{H}^{\prime}$ (a) and third harmonic 
	amplitude $|T_{H3}|$ (b) measured on an optimally doped 
	Bi$_{2}$Sr$_{2}$CaCu$_{2}$O$_{8+\delta}$ single crystal ($T_{c} = 
	90.0$ K), irradiated with 5.8 GeV Pb ions to a matching field 
	$B_{\phi} = 0.5 $ T. The measurement frequency was 7.753 Hz and the 
	ac field amplitude $h_{ac} = 6$ Oe. DC field values are as indicated. }
	\label{fig:THTH3}
	\end{figure}

\noindent	pinning of 
vortices yields the rather unlikely result that the entropy contribution 
to the free energy rapidly \em decreases \rm as function of temperature.
	
\subsection{Relation with the phase diagram}
\label{section:irr}

\begin{figure}
	\centerline{\epsfxsize 8cm \epsfbox{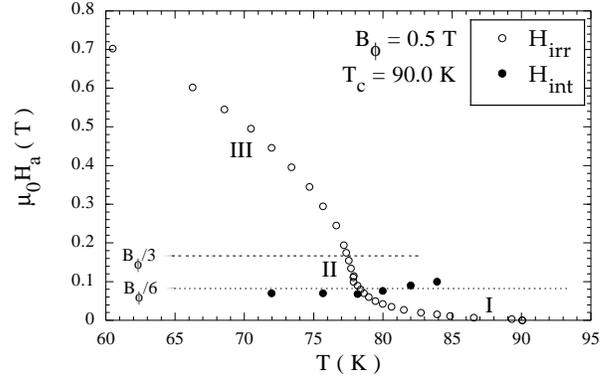}}
	\caption{Irreversibility field $H_{irr}(T)$, at which a third 
	harmonic ac response $|T_{H3}|$ can first be measured upon cooling 
	the crystal (the same as in Fig.~\protect\ref{fig:THTH3}) in a magnetic 
	field ($\circ$). The roman numerals indicate the field regimes 
	described in the text. Also indicated are the fields $H_{int}(T)$ determined from 
	reversible magnetization measurements on the same crystal ($\bullet$).}
	\label{fig:Hint}
	\end{figure}

In order to correlate the vortex arrangement on the columnar tracks
as revealed by the reversible magnetization measurements with the phase 
diagram, we have performed measurements of the ac transmittivity as 
described in section~\ref{section:transmittivity}. Typical results 
for the in-phase fundamental and third harmonic amplitude 
are shown in Fig.~\ref{fig:THTH3} for optimally doped 
Bi$_{2}$Sr$_{2}$CaCu$_{2}$O$_{8+\delta}$ ($T_{c} = 90.0$ K) with 
$B_{\phi} = 0.5$ T.\cite{0.5T} The corresponding fields $H_{irr}(T)$ 
at which a third harmonic response is first detected are plotted in 
Fig.~\ref{fig:Hint}. Note  that due to the different 
effective measurement frequencies, $f = 7.753$ Hz in the ac shielding 
experiment versus $f \sim 10^{-2}$ Hz for magnetization measurements 
using the SQUID, the   $H_{irr}$--values plotted in the Figure lie much above the 
apparent $H_{irr}$--values extracted from the closing of the magnetic 
hysteresis loops or merger of zero--field--cooled and field--cooled 
data obtained with a SQUID magnetometer.\cite{Zech96,Sato97,Kosugi97,Chikumoto98}
The same is indicated in Fig.~\ref{fig:Mrev-T}(a), which shows the 
position of $H_{irr}$ determined from transmittivity measurements on 
the magnetization curve. The magnetization data show a break at 
$H_{irr}$, reflecting the inadequacy of moving

\noindent	sample SQUID magnetometry for 
measurements in the regime $H_{a} \lesssim H_{irr}$. Another reason 
for the much lower $H_{irr}(T)$ values of 
Ref.~\onlinecite{Kosugi97,Chikumoto98} is the fact that those authors 
used lightly underdoped crystals. While $T_{c}$ can be similar for 
underdoped and overdoped samples, it was recently shown that the 
irreversibility lines for the irradiated material vary widely as 
function of oxygen content.\cite{vdBeek99}

\begin{figure}
	\centerline{\epsfxsize 8cm \epsfbox{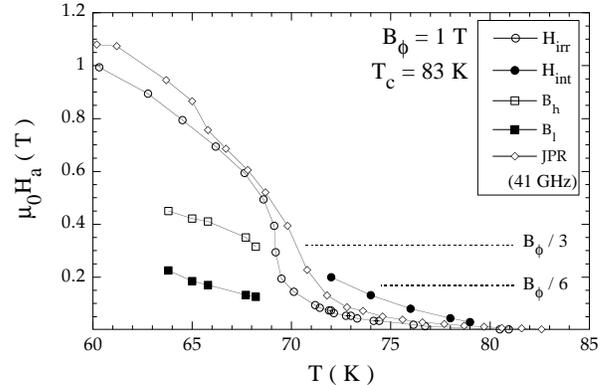}}
    \caption{Irreversibility field $H_{irr}(T)$ determined for lightly 
    overdoped Bi$_{2}$Sr$_{2}$CaCu$_{2}$O$_{8+\delta}$,\protect\cite{1T} with 
    $B_{\phi} = 1$ T ($\circ$). The crystal was chosen so as to have $T_{c} = 
    83.1$ K comparable to the sample used to obtain the JPR peak data 
    of Ref.~\protect\onlinecite{Sato97} (also shown ($\diamond$)), as 
    well as to the crystal of Ref.~\protect\onlinecite{Morozov98}. The
    characteristic fields $B_{l}$ (\protect\rule{1.5mm}{1.5mm}) at which the 
    $c$--axis conductivity starts to increase, and $B_{h}$ 
    ( \protect\raisebox{1mm}{\protect\fbox{\protect\small }} ) at which it starts 
    its final decrease \protect\cite{Morozov98} are displayed as 
    filled and open squares, respectively. The interaction fields $H_{int}(T)$ 
	determined from reversible magnetization measurements are drawn as 
	filled circles ($\bullet$).}
	\label{fig:phasediag}
	\end{figure}

The irreversibility field diplays three distinct regimes as function 
of temperature. Upon lowering temperature from $T_{c}$, $H_{irr}(T)$ 
first increases exponentially (I);\cite{Hardy93,vdBeek95I,Zech96,vdBeek95II} 
at $T \simeq 77$ K, $H_{irr}$ increases sharply (II), before bending over 
to an approximately linear temperature dependence below 70 K (III). 
The same behavior is found for other matching fields, see {\em e.g.} 
Fig.~\ref{fig:phasediag} for $B_{\phi} = 1$ T; for

	\begin{figure}[t]
	\centerline{\epsfxsize 8cm \epsfbox{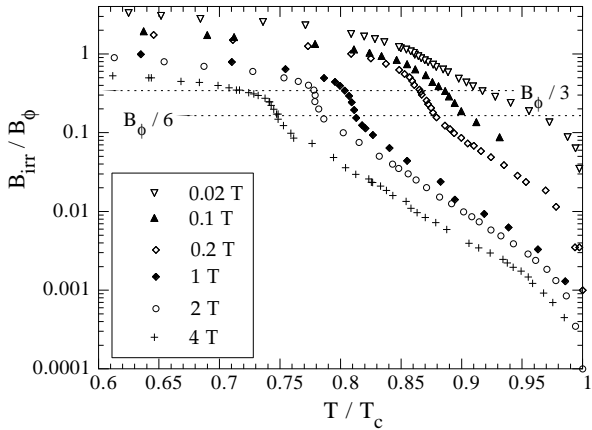}}
	\caption{Dependence of $H_{irr}(T)$ on matching field. Displayed are 
	the values $\mu_{0}H_{irr}(T) / B_{\phi}$ for a series of crystals 
	grown at the University of Tokyo, irradiated with 
	5.8 GeV Pb ions to matching fields between 0.02 T and 4 T; data are 
	plotted as function of $T/T_{c}$, where $T_{c}$ is the value \em after 
	\rm irradiation ({\em e.g.} $T_{c} = 86.2$ K for $B_{\phi} = 2$ T).}
	\label{fig:IRL-Bfi}
	\end{figure}

	\noindent   comparison, the same Figure shows 
characteristic fields extracted  from $c$-axis conductivity and JPR measurements.
It is seen that the different regimes of the irreversibility 
line $H_{irr}(T)$ are faithfully reproduced by the position of the 
JPR peak;\cite{Sato97} for suitably chosen measurement 
frequency (here, 41 GHz) the two lines lie very close together,
suggesting that the onset of linear vortex diffusion  (and 
hence, the disappearance of long-range $ab$-plane phase coherence) indicated by 
$H_{irr}(T)$ is intimately related to the drop in $c$-axis phase 
coherence expressed by the JPR peak.
	
Also plotted in either Figure are the $H_{int}$--data for the same 
crystals. For the optimally doped crystal ($B_{\phi} = 0.5$ T) of 
Fig~\ref{fig:Hint}, it was possible to obtain $H_{int}$--values directly from the 
reversible magnetization for fields \em below  \rm
the irreversibility field by exploiting the torque data of 
Ref.~\onlinecite{Drost98}. The reversible magnetization curves in 
this case were measured with the magnetic field applied at a 
substantial angle with respect to the $c$-axis and the column direction. 
It was shown previously that in such a configuration, the reversible 
magnetization depends only on the field component $H_{z}$ parallel to the 
$c$-axis; within experimental error, the magnetization curve $M_{rev}(H_{z})$ is the same 
regardless of the direction of the total magnetic field. Simultaneously, 
applying the field at an angle lowers the irreversibility field $H_{irr}$ 
to a much lower value than when the field is aligned with the columnar 
defects.\cite{vdBeek95I}

It appears that the field $H_{int}$ intercepts the irreversibility 
line exactly at the point $\mu_{0}H_{a} \approx \frac{1}{6}B_{\phi}$ 
where $H_{irr}$ starts its steep rise (Fig~\ref{fig:Hint}). 
This strongly suggests that the reduction of
vortex mobility causing the increase in	 $H_{irr}(T)$ is the 
consequence of intraplane vortex repulsion and the reduction of 
the number of sites available to a given vortex.  The $H_{int}$ data 
nicely coincide with the recoupling field obtained in 
Refs.~\onlinecite{Sato97,Kosugi97}. For $B_{\phi} = 1$ and 2 T, 
reliable $H_{int}$-data could not be obtained in the temperature 
interval spanning the irreversibility line. The data in 
Fig.~\ref{fig:phasediag} show that for temperatures well in excess of 
$H_{irr}$, $H_{int}$ decreases with increasing $T$. The  
temperature dependence of the interaction field, reminiscent of that 
of the high--field part of the irreversibility line, seems to extrapolate 
to the field $B_{h}$ where the $c$--axis conductivity starts its 
ultimate decrease.\cite{Morozov98} Note that $H_{int}$ never increases 
beyond the field $\mu_{0}H_{a} \approx \frac{1}{6}B_{\phi}$ at which 
the recoupling transition is expected. This particular fraction of the 
matching field, $\frac{1}{6}B_{\phi}$, seems to be quite robust. Figure~\ref{fig:IRL-Bfi}, which 
displays the evolution of $H_{irr}(T)$ with $B_{\phi}$, shows that 
for 0.1 T $ \leq B_{\phi} \leq 4$ T the steep rise of the irreversibility 
line always starts near $\mu_{0}H_{a} \approx \frac{1}{6}B_{\phi}$. 
This is in contradiction to numerical work \cite{Sugano98} and previous magnetization 
studies \cite{Chikumoto98} which reported significant features in 
the phase diagram at $\mu_{0}H_{a} \approx \frac{1}{3}B_{\phi}$. 
Figure~\ref{fig:IRL-Bfi} shows that, whereas for the values of the 
matching field $B_{\phi} = 1 $ and 2 T, most commonly investigated in 
the literature,\cite{Sato97,Kosugi97,Chikumoto98,Sugano98} 
the fraction $B_{\phi}/3$ coincides with the 
crossover from regime (II) to regime (III) in $H_{irr}(T)$, this is by 
no means true for other values of $B_{\phi}$. The results also show 
that the reported scaling of the irreversibility line with $B_{\phi}$ 
reported in Ref.~\onlinecite{Hardy93} is by no means to be taken 
strictly.

\section{Discussion}

A number of significant results emerge from the above. The most 
notable is that the introduction of amorphous columnar defects into the 
matrix of the layered superconductor decreases the free energy in the 
London regime not only through vortex pinning, but also 
through an additional configurational entropy, which appears when a 
sufficiently large number of extra sites become available 
to the pancake vortices. This extra contribution can be 
comparable to or larger than the pinning energy. Previously presented 
values of the pinning energy extracted from the reversible 
magnetization of crystals with relatively high matching fields 
\cite{Wahl95,vdBeek96II,QiangLi96,Hardy98} should thus be considered 
with reserve. More accurate measurements of the pinning energy for 
crystals with lower $B_{\phi}$ were recently presented in 
Ref.~\onlinecite{Drost99}. The pinning energies obtained in this 
paper are in agreement with those of Ref.\onlinecite{Drost99} and 
with the vortex-core pinning model, $U_{0} \approx 
\varepsilon_{0}(c_{0}/2 \xi)^{2}$.\cite{Nelson92} This means that 
even in the vortex liquid state, vortices are, certainly in the 
regime $B \ll B_{\phi}$, strongly bound to the 
columnar defects. Thermally induced vortex wandering from the tracks 
would have resulted in a pinning energy contribution $U_{0}/\Phi_{0}$
that is exponentially small with respect to the total 
magnetization.\cite{Koshelev96} The neglect of the entropy contribution
also accounts for the discrepancy between 
experimental results and the numerical simulations of
Ref.~\onlinecite{Wengel98}. The  curves calculated in Ref.~\onlinecite{Wengel98}
are close to what would be expected in the (near--) absence of an entropy contribution 
(see for example data at $B_{\phi} = 0.2$ T). Conversely, the use of 
Eq.~(\ref{eq:Bulaevskiic1}) overestimates the entropy contribution 
in many cases and yields pinning energies 

 \begin{figure}
	\centerline{\epsfxsize 8cm \epsfbox{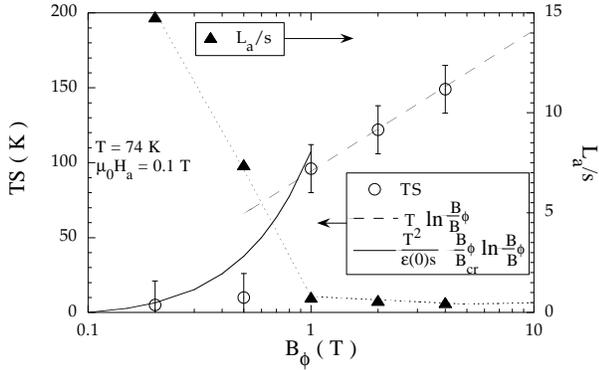}}
    \caption{Entropy contribution to the free energy of heavy-ion 
    irradiated lightly overdoped Bi$_{2}$Sr$_{2}$CaCu$_{2}$O$_{8+\delta}$ ($T_{c} = 83$ 
    K) as function of matching field ($\circ$), for $T = 74 $ K and applied field $\mu_{0}H_{a} = 
    0.1$ T. The entropy rises steeply between $B_{\phi} = 0.5$ and 1 
    T, from where it increases logarithmically as function of 
    $B_{\phi}$. This behavior is in qualitative agreement with the 
    dissociation of individual vortex lines into stacks of length 
    $L_{a}$ at low defect density (few available sites per vortex), 
    for which the prediction $TS \propto B_{\phi}\ln B_{\phi}$ is 
    indicated by the curved full line; at higher matching fields, the 
    length of the typical stack has decreased to the layer spacing 
    $s$ --- the dependence then becomes $TS \propto \ln B_{\phi}$ 
    corresponding to the configurational entropy associated with the 
    possibility for each pancake vortex to occupy different columns 
    (dashed straight line). The corresponding values of $L_{a}$ are 
    indicated by the full triangles (the dotted line is a guide to the eye).
	The parameter values used are $\varepsilon_{0}s = 1500$ K,
	$B_{cr}=0.4$ T,\protect\cite{vdBeek91} \em i.e. \rm $\gamma = 
    55$.\protect\cite{Farrell90}}
	\label{fig:TS(Bfi)}
	\end{figure}

\noindent that are manifestly too 
small.\cite{Bulaevskii96II}

The measurement of the reversible magnetization as function of 
column density allows a direct determination of the extra entropic 
contribution $TS$ (Fig.~\ref{fig:Mrev-Bfi}), and its dependence on field and temperature.
The mechanisms by which the free energy is lowered becomes apparent 
if one considers the rather peculiar dependence of $TS$ onthe density of sites at 
constant field, illustrated in Fig.~\ref{fig:TS(Bfi)}. At low 
$B_{\phi}$, $TS$ remains small in comparison with other terms 
contributing to the free energy. This can also be inferred directly 
from the near--equality of the low-- and high field limits of the 
logarithmic slope of  the reversible magnetization for $B_{\phi} \lesssim 0.5$ 
T.\cite{Drost98,Drost99} At high matching field, $TS$ follows a 
logarithmic dependence on ion dose. We interpret this behavior as 
follows. At low irradiation doses, the increase in entropy with increasing 
columnar defect density arises from 
the dissociation of vortex lines into shorter and shorter stacks of length $L_{a}$ 
localized on different columns; the entropic part of the free energy 
(in the vortex liquid phase) then obeys Eq.~(\ref{eq:correlated}), with 
$\beta \sim \frac{1}{2}$. As a result, $TS$ increases as $B_{\phi}\ln B_{\phi}$, which 
explains its rapid rise between $B_{\phi} = 0.5$ T and 1 T and the 
quite important change in the low--field magnetization between these 
matching fields. Once the length of a typical stack has decreased to a 
single pancake vortex, this mechanism can no longer be operative. The 
entropy now increases as $\ln B_{\phi}$, corresponding to the 
possibility for each single pancake to occupy different columnar defects. 
The evolution of the entropy with increasing matching field 
can be described by the optimisation of the total free energy gain, 
obtained from the balance between the entropy gain and the loss in intervortex 
interaction energy. At small $B_{\phi}$, there are few nearlying 
columnar defect sites per vortex line --- the access to farlying 
sites is prohibited by intervortex repulsion. Hence, vortices are on 
average well--aligned on the same site. For increasing $B_{\phi}$, 
more and more nearlying defect sites become available, and the free 
energy can be lowered through the dissociation of vortex lines, 
first into stacks of pancakes located on adjacent columns, and finally into 
single pancakes wandering between available column sites.

The same explains the field dependence of $TS$ (Figs.\ref{fig:Mrev-T} and
\ref{fig:Mrev-Bfi}). As more and more vortices are added, fewer sites are 
available per vortex. First, the entropy contribution drops (logarithmically) 
because fewer defect sites are available to each pancake. At higher fields, pancakes 
belonging to the same vortex line become aligned on the same site, be 
this on a columnar defect or on an interstitial site. This leads to a 
reduction in vortex mobility, and to the increase of the $c$-axis critical 
current measured by JPR\cite{Kosugi97,Bulaevskii98} and of the 
$c$-axis conductivity.\cite{Morozov98,Morozov99,remarkMorozov} 

It is to be stressed that in the presence of columns 
the low--field entropy contribution to the free 
energy  is much \em larger \rm than that in 
unirradiated crystals, at odds with the prediction of 
Eqs.~(\ref{eq:Bulaevskii},\ref{eq:Bulaevskiic1}).\cite{Bulaevskii96II}
A qualitative explanation for this is that, whereas in unirradiated crystals 
one has one local energy minimum per vortex lattice unit cell, the presence of a columnar 
defect--density $n_{d} > B / \Phi_{0}$ amounts to the presence of 
several energy minima per cell, permitting the ``roughening'' of vortex 
lines (\em i.e. \rm their spreading over different columnar defects). 
Entropy decreases as function of $B$ until at $B \sim \frac{1}{2}B_{\phi}$ 
it attains a value close to that measured in 
unirradiated samples, corresponding to approximately one site per 
vortex. From this field onwards, it becomes difficult to obtain a 
reliable estimate of the alignment length $L_{a}$ from reversible 
magnetization. This is because of the ``error bar'' introduced by the
large value of the Ginzburg number $Gi$. We have seen in Section 
\ref{section:entropy} there is an uncertainty of the order $0.1 
k_{B}T/\Phi_{0}s$ when one determines the relative importance of the different 
terms contributing to $\partial M_{rev} / \partial \ln B$. In 
particular, one cannot estimate the entropy contribution to better 
than $0.1 k_{B}T/\Phi_{0}s$, corresponding to a lower bound $L_{a} = 
10 s$. The same holds for unirradiated crystals. 
Whereas, in the irradiated superconductor,  the dissociation of vortex lines into stacks of pancakes 
(within a unit cell) certainly lowers interplane correlations of the phase 
of the superconducting order parameter, it does not necessarily imply the 
loss of long--range phase coherence. This is likely destroyed near the 
irreversibility line, as linear activated motion of (pancake) vortices becomes possible.

The questions of pancake vortex alignment, phase coherence, and the 
recoupling transition reported in Refs.~\onlinecite{Sato97,Kosugi97,Chikumoto98} 
are intimately related. From reversible magnetization, we have determined the 
field $H_{int}$ at which intervortex repulsion determines the vortex pancake arrangement in 
the presence of columnar defects. This is in very good 
agreement with the recoupling field obtained in 
Refs.~\onlinecite{Sato97,Kosugi97}, and coincides with the rapid rise 
of $H_{irr}(T)$. It seems that the same intraplane vortex repulsion 
responsible for vortex alignment also decreases the vortex mobility, 
expressed by the peak in the critical current density measured in 
Refs.~\onlinecite{Chikumoto98} and \onlinecite{Konczykowski94I}, and by 
the increase in $H_{irr}(T)$ into regime (II) at $\frac{1}{6}B_{\phi}$. As for the 
increase in vortex mobility and the decrease of $c$--axis 
conductivity at the crossover of $H_{irr}$ from regime (II) to the 
high--field region (III), it is likely that it corresponds to the 
field at which the number of free vortices becomes comparable to, or exceeds 
the number of vortices trapped on a columnar defect. This field can be 
roughly estimated as that at which $U_{p} + TS$ 
(Figs.~\ref{fig:Mrev-T}(b) and \ref{fig:Mrev-Bfi}(b)) has decreased to half 
its value at $H_{int}$. For $B_{\phi} \gtrsim 1$ T, it amounts to 
roughly the value $\frac{1}{3}B_{\phi}$ cited in 
Ref.~\onlinecite{Chikumoto98}; for lower matching fields it lies closer 
to $B_{\phi}$. In all cases, it lies close to the crossover in question. 
Thus, we identify regime (III) of $H_{irr}(T)$ as that in which the 
number of free vortices exceeds the number of trapped ones, and in 
which vortex mobility is determined by plastic vortex 
creep.\cite{Nelson92}

\section{Conclusion}

Reversible magnetization measurements have been analyzed in order to 
determine the relative importance of pinning energy and entropy 
contributions to the free energy of heavy--ion irradiated 
Bi$_{2}$Sr$_{2}$CaCu$_{2}$O$_{8+\delta}$ single crystals. We have 
found that in unirradiated crystals the entropy contribution to the 
free energy in the London regime is relatively minor. This is also 
true in irradiated crystals with small matching fields $B_{\phi} 
\lesssim 0.5$ T, but not for irradiated crystals with large $B_{\phi} 
\gtrsim 1$ T and fields $B \ll B_{\phi}$. Then, the configurational entropy 
contribution to the free energy is greatly enhanced with respect to 
its value in unirradiated crystals, and can easily exceed the 
pinning energy of a vortex on a columnar defect by more than a factor 2. 
The large entropy contribution at high 
$B_{\phi}$ is a consequence of the dissociation of vortex lines into 
individual pancake vortices localized on different columnar defect 
sites. As $B/B_{\phi}$ is increased, intervortex 
repulsion causes the pancakes belonging to the same stack to align 
onto the same columnar defect. It was possible to obtain a 
lower bound $L_{a} \sim 10 s$ on the length over which pancakes are 
aligned at $B \approx B_{\phi}$. Vortex interactions become important 
at the field $H_{int} \sim \frac{1}{6} B_{\phi}$; at this field,  
which delimits the regime of pinning of individual vortices, and which 
corresponds to the recoupling field measured by Josephson Plasma Resonance, the 
irreversibility field sharply increases.  Above $H_{int}$, the repulsive 
interaction between vortices causes both the vortex mobility to 
decrease and pancake alignement to increase. At higher fields $\gtrsim 
\frac{1}{3}B_{\phi}$, free vortices outnumber those that are trapped on a columnar defect. 
This causes the decrease of $c$-axis resistivity and a second crossover 
of the irreversibility field, to a regime where it is determined by plastic 
creep.

We gratefully acknowledge J. Blatter, L.N. Bulaevskii, A. Buzdin, M.V. Feigel'man, A.E. 
Koshelev, P. LeDoussal, Y. Matsuda, T. Tamegai, and V.M. Vinokur for most stimulating 
discussions; We thank A.A. Menovsky of the Netherlands Organisation for Fundamental 
Research on Matter  - Amsterdam--Leiden Metals Research Collaboration 
(F.O.M. - A.L.M.O.S.), as well as N. Motohira (University of Tokyo) for providing the 
Bi$_{2}$Sr$_{2}$CaCu$_{2}$O$_{8+\delta}$ single crystals.

\newpage


\begin{thebibliography}{99}

\bibitem{Hardy92} V. Hardy, J. Provost, D. Groult, M. Hervieu, B. Raveau, 
S. Dur\v{c}ok, E. Pollert, J. C. Frison, J.P. Chaminade, and M. Pouchard, 
Physica C{\bf 191}, 85 (1992).

\bibitem{Konczykowski93} M. Koncykowski, Y. Yeshurun, L. Klein, E.R. 
Yacoby, N. Chikumoto, V.M. Vinokur, and M.V. Feigel'man, J. Alloys Comp. 
{\bf 195}, 407 (1993).

\bibitem{Hardy93} V. Hardy, Ch. Simon, J. Provost, and D. Groult, 
Physica C {\bf 205}, 371 (1993).

\bibitem{Klein94I} L. Klein, E.R. Yacoby, Y. Yeshurun, M. Konczykowski, 
and K. Kishio, Phys. Rev B {\bf 48}, 4403 (1994).

\bibitem{Konczykowski95} M. Konczykowski, N. Chikumoto, V.M. Vinokur, and 
M.V. Feigel'man, Phys. Rev. B {\bf 51}, 3957 (1995).

\bibitem{vdBeek95I} C.J. van der Beek, M. Konczykowski, V.M. Vinokur, T.W. 
Li, P.H. Kes, and G.W. Crabtree, Phys. Rev. Lett. {\bf 74}, 1214 (1995).

\bibitem{Zech95}  D. Zech, S.L. Lee, H. Keller, J. Blatter, B. 
Janossy, P.H. Kes, and T.W. Li, Phys. Rev. B {\bf 52},  6913 (1995).

\bibitem{Zech96} D. Zech, S.L. Lee, H. Keller, J. Blatter, P.H. Kes, and 
T.W. Li, Phys. Rev. B {\bf 54},  6129 (1996).

\bibitem{Seow96}W.S. Seow, R.A. Doyle, A.M. Campbell, G. Balakrishnan, K.
Kadowaki, and G. Wirth, Phys. Rev. B {\bf 53}, 14611 (1996).

\bibitem{Doyle96} R. Doyle, W.S. Seow, Y. Yan, A.M. Campbell, T. 
Mochiku, K. Kadowaki, and G. Wirth, Phys Rev. Lett., Phys. Rev. Lett. {\bf 77}, 1155 
(1996).

\bibitem{Sato97} M. Sato, T. Shibauchi, S. Ooi, T. Tamegai, and M. 
Konczykowski, Phys. Rev. Lett. {\bf 79}, 3759 (1997).

\bibitem{Kosugi97} M. Kosugi, Y. Matsuda, M.B. Gaifullin, L.N. Bulaevskii, 
N. Chikumoto, M. Konczykowski, J. Shimoyama, K. Kishio, K. Hirata, and K. 
Kumagai, Phys. Rev. Lett. {\bf 79}, 3763 (1997).

\bibitem{Chikumoto98} N. Chikumoto, M. Kosugi, Y. Matsuda, M. Konczykowski, 
and K. Kishio, Phys. Rev. B {\bf 57}, 14507 (1998).

\bibitem{Wahl95} A. Wahl, V. Hardy, J. Provost, C. Simon, and A. Buzdin, 
Physica C {\bf 250}, 163 (1995).

\bibitem{vdBeek96II} C.J. van der Beek, M. Konczykowski, T.W. Li, P.H. 
Kes, and W. Benoit, Phys. Rev. B. {\bf 54}, R792 (1996).

\bibitem{QiangLi96}Qiang Li, Y. Fukumoto, Y.M. Zhu, M. Suenaga, T. Kaneko, 
K. Sato, and Ch. Simon, Phys. Rev B {\bf 54}, R788,1996.

\bibitem{Drost98} R.J. Drost, C.J. van der Beek, J.A. Heijn, M. Konczykowski, 
and P.H. Kes, Phys. Rev. B {\bf 58}, R615 (1998).

\bibitem{Drost99} R.J. Drost, C.J. van der Beek, H.W. Zandbergen, M. Konczykowski,
A.A. Menovsky, and P.H. Kes, Phys. Rev. B {\bf 60},  (1998).

\bibitem{Kes90} P.H. Kes, J. Aarts, V.M. Vinokur, and C.J. van der Beek,  
Phys. Rev. Lett. {\bf 64}, 1064 (1990).

\bibitem{Clem91I} J.R.Clem, Phys. Rev. B {\bf 43}, 7837 (1991).

\bibitem{Nelson92} D.R. Nelson and V.M. Vinokur, Phys. Rev. Lett. 
{\bf 68}, 2398 (1992); {\em ibid.}, Phys. Rev. B {\bf 48}, 13060.

\bibitem{Iye91} Y. Iye, T. Tamegai, and S. Nakamura, Physica C {\bf 174}, 
227 (1991).

\bibitem{Raffy91} H. Raffy, S. Labdi, O. Laborde, and P. Monceau, Phys. Rev. 
Lett. {\bf 66}, 2515 (1991).

\bibitem{Martinez92} J.C. Martinez, S. Brongersma, A. Koshelev, B. Ivlev, 
P.H. Kes, R.P. Griessen, D.G. de Groot, Z. Tarnavski, and A.A. Menovsky, Phys. Rev. 
Lett. {\bf 69}, 2276 (1992).

\bibitem{Matsuda95} Y. Matsuda, M.B. Gaifullin, K. Kumagai, K. Kadowaki, 
and T. Mochiku, Phys. Rev. Lett. {\bf 75}, 4512 (1995).

\bibitem{Morozov98} N. Morozov, M.P. Maley, L.N. Bulaevskii, and J. 
Sarrao, Phys. Rev. B {\bf 57}, R8146 (1998).

\bibitem{Morozov99} N. Morozov, M.P. Maley, L.N. Bulaevskii, V. 
Thorsm\o lle, A.E. Koshelev, A. Petrean, and W.K. Kwok, Phys. Rev. 
Lett. {\bf 82},  (1999).

\bibitem{Hanaguri97} T. Hanaguri, Y. Tsuchiya, S. Sakamoto, A. Maeda, and 
D.G. Steel, Phys. Rev. Lett. {\bf 78}, 3177 (1997).

\bibitem{Kosugi98} M. Kosugi, Y. Matsuda, M.B. Gaifullin, L.N. Bulaevskii, 
N. Chikumoto, M. Konczykowski, J. Shimoyama, K. Kishio, and K. Hirata, 
Phys. Rev. B {\bf   },  (1998).

\bibitem{Bulaevskii98} L.N. Bulaevskii, M.P. Maley, and V.M. Vinokur, 
Phys. Rev. B {\bf 57}, R5626 (1998).

\bibitem{Wengel98} C. Wengel and U.C. T\"{a}uber, Phys. Rev. Lett. 
{\bf 78}, 4845 (1997); C. Wengel and U.C. T\"{a}uber, Phys. Rev. B
{\bf 58},  6565 (1998).

\bibitem{Hardy98} V. Hardy, S. H\'{e}bert, M. Hervieu, Ch. Simon, J. 
Provost, A. Wahl, and A. Ruyter, Phys. Rev. B {\bf 58}, 15218 (1998).

\bibitem{TWLi94} T.W. Li, P.H. Kes, N.T. Hien, J.J.M. Franse, and A.A. 
Menovsky, J. Cryst. Growth {\bf 135}, 481 (1994).

\bibitem{Motohira} N. Motohira, K. Kuwahara, T. Hasegawa, K. Kishio, 
and K. Kitazawa, J. Ceram. Soc. Jpn. Int. Ed. {\bf 97}, 994 (1989).

\bibitem{Gilchrist93} J. Gilchrist and M. Konczykowski, Physica (Amsterdam)
C {\bf 212}, 43 (1993).

\bibitem{vdBeek95II} C.J. van der Beek, M. Konczykowski, V.M. Vinokur, 
T.W. Li, P.H. Kes, and G.W. Crabtree, Phys. Rev. B {\bf 51}, (1995).

\bibitem{Pradhan95} A.K. Pradhan, S.B. Roy, P. Chaddah, C. Chen, and B.M. Wanklyn, Phys. Rev. B
{\bf 52}, 6215 (1995).

\bibitem{VB5} This is the same crystal as that used in 
Ref.\protect\onlinecite{vdBeek96II}, Fig. 1.

\bibitem{Clem} Z. Hao and J.R. Clem, Phys. Rev. Lett. {\bf 67}, 2371 
(1991).

\bibitem{Koshelev94II} A.E. Koshelev, Phys. Rev. B {\bf 50}, 506 (1994).

\bibitem{TWLi96} T.W. Li {\em et al.},  
Physica (Amsterdam) C {\bf 257}, 179 (1996).

\bibitem{Kogan93} V.G. Kogan, M. Ledvij, A. Yu. Simonov, J.H. Cho, and 
D.C. Johnston, Phys. Rev. Lett. {\bf 70}, 1870 (1993).

\bibitem{Bulaevskii92} L.N. Bulaevskii, M. Ledvij, and V.G. Kogan, Phys. 
Rev. Lett. {\bf 68}, 3773 (1992).

\bibitem{Bulaevskii96II} L.N. Bulaevskii, V.M. Vinokur, and M.P. Maley, 
Phys. Rev. Lett. {\bf 77}, 936 (1996).

\bibitem{Mrktchyan71} G.S. Mkrtchyan and V.V. Shmidt, Zh. Eksp. Teor. Fiz. 
{\bf 61}, 367 (1971) [Sov. Phys. JETP {\bf 34}, 195 (1972)].

\bibitem{Buzdin94} A.I. Buzdin and D. Feinberg, Physica C {\bf 235}--{\bf 
240}, 2755 (1994).

\bibitem{Ikeda95} R. Ikeda and T. Tsuneto, J. Phys. Soc. Jap. {\bf 
60}, 1337 (1991); R. Ikeda, J. Phys. Soc. Jap. {\bf 64}, 1683 (1995).

\bibitem{Tesanovic92}Z. Te\v{s}anovi\'{c} {\it et al.}, Phys. Rev. Lett. 
{\bf 69}, 3563 (1992).

\bibitem{Kes91} P.H. Kes, C.J. van der Beek, M.P. Maley, M.E. McHenry, 
D.A. Huse, M.J.V. Menken, and A.A. Menovsky, Phys. Rev. Lett. {\bf 67}, 2383 (1991)

\bibitem{QiangLi93} Qiang Li, K. Shibutani, M. Suenaga, I. Shigaki, and  R. 
Ogawa, Phys. Rev. B {\bf 48}, 9877 (1993).

\bibitem{Blatter} We are indebted to Prof. Dr. J. Blatter for this 
estimate. A similar estimate can be found in 
Ref.~\protect\onlinecite{Koshelev96}. However, the latter is 
inapplicable because it describes the situation where pancakes vortices 
are weakly localized on the columnar defects. Experimentall, we find 
that thermal wandering of vortices trapped on a columnar 
defect can be neglected.

\bibitem{Koshelev96} A.E. Koshelev, P. LeDoussal, and V.M. Vinokur, Phys. 
Rev. B {\bf 55}, R8855 (1996).

\bibitem{0.2T} For the lowest matching fields ({\em e.g.} 0.2 T) 
difficulties arise again because the low field limit of $M_{rev}$ 
lies entirely within the irreversible regime.

\bibitem{0.5T} This is the same sample as in 
Refs.~\protect\onlinecite{Drost98,Drost99}.

\bibitem{vdBeek99} C.J. van der Beek, M. Konczykowski, A.V. Samoilov, N. 
Chikumoto, S. Bouffard, and M.V. Feigel'man, preprint (1999);

\bibitem{1T} This is the same crystal as in Ref.
\protect\onlinecite{vdBeek96II}, Figs.~2 and 3.

\bibitem{Sugano98} R. Sugano, T. Onogi, K. Hirata, and M. Tachiki, Phys. 
Rev. Lett. {\bf 80}, 2925 (1998).


\bibitem{vdBeek91} C.J. van der Beek and P.H. Kes, Phys. Rev. B {\bf 43}, 
13032 (1991).

\bibitem{Farrell90} D.E. Farrell, S. Bonham, J. Foster, Y.C. Chang, P.Z. 
Jiang, K.G. Vandervoort, D.J. Lam, and V. G. Kogan, Phys. Rev. Lett. {\bf 63}, 782 
(1990).

\bibitem{remarkMorozov} Although Ref.~\protect\onlinecite{Morozov99} 
states that vortex interactions determine the positional 
correlations in the vortex liquid, it cannot make statements about 
which kind of vortex interaction: intralayer repulsion or interlayer 
attraction.

\bibitem{Konczykowski94I} M. Konczykowski, Physica C {\bf 235}-{\bf 240}, 
197 (1994); M. Konczykowski, N. Chikumoto, V. Vinokur, M. 
Feigel'man, Physica C {\bf 235}-{\bf 240}, 2845 (1994).

\end{thebibliography}
\end{document}